 \documentclass[fleqn,twoside]{article}
 \usepackage{espcrc2}
 
\usepackage{graphicx}
\usepackage{graphics}
 
\newcommand{\AmS}{{\protect\the\textfont2
  A\kern-.1667em\lower.5ex\hbox{M}\kern-.125emS}}
 
\usepackage{color}
 
\title{Empirical Baye's Method and
Tests in Very Light Quark Range from
The Overlap Lattice QCD} 

\author{S.J. Dong\address[UK]{Dept. of Physics and Astronomy, University 
of Kentucky, Lexington, KY 40506},
T. Draper\addressmark[UK], I. Horvath\addressmark[UK],
F.X. Lee\address{Center for Nuclear Studies, Dept. of Physics, George Washington University,
Washington, DC 20052}\address{Jefferson Lab, 12000 Jefferson Avenue, 
Newport News, 
VA 23606}, Nilmani Mathur\addressmark[UK]
and J.B. Zhang\address{CSSM and Dept. of Physics and Math. Physics,
University of Adelaide, Adelaide, SA 5005, Australia}}
\begin{document}
\begin{abstract}
Based on Bayesian theorem an empirical Baye's method is discussed. A
programming chart for mass spectrum fitting is suggested. A weakly constrained
way for getting priors to solve the chiral log data fitting singularity
is tested.
\end{abstract}
\maketitle
\section{Introduction}
Data analysis is an important procedure for most numerical projects and
experiments. In lattice QCD, the theory we fit to hadronic two point
correlation function
is:
\begin{eqnarray}
G(t,A,m) &=& \sum_{i=1}^{\infty} A_i e^{-m_i (t-t_0)} 
\end{eqnarray}
However, the general minimum $\chi^2$ procedure does not work
for such a full physics hypothesis as the procedure is singular. What
we used to do is to truncate both data set and theory. We might fit only
the large-$t$ behavior with the lowest mass particle.

Numerically, the minimizing procedure is equivalent to a linear equation.   
Suppose ${\bf \rho}=\{A_i,m_i\}$ is an unknown parameter vector
\begin{eqnarray}
&&{\rm minimize:}~~~\mathcal{A}({\bf \rho}) \nonumber \\
&\Rightarrow & \mid {\bf A\cdot \rho} - {\bf c}\mid = 0
\end{eqnarray} 

 A singularity can occur due to the fact that matrix ${\bf A}$ 
has is degenerate or has zero eigenvalues.
A possible solution to the singularity is to add another
matrix ${\bf B}$.
Suppose there is another minimizing procedure $\mathcal{B}({\bf \rho})$,
\begin{eqnarray}
&&\mathcal{C}(\rho) = \mathcal{A}(\rho)+\lambda \mathcal{B}(\rho) \\
&\Rightarrow& \mid {\bf C\cdot \rho} - {\bf c}^\prime\mid = 0 \nonumber
\end{eqnarray}
We can minimize $\mathcal{C}(\rho)$
instead of $\mathcal{A}(\rho)$.

\section{Bayesian Theory}
Bayesian statistics provides a useful way to offer such an additional
minimizing procedure. The parameter vector $\rho$ describes the hypothesis
of Eq. (1). To get $\rho$ from the measured data set $D$ is a numerical
{\it inverse} problem. The Bayesian theorem tells us that \cite{book}:
 a single good inverse
${\bf \rho}$, is to maximize the probability
\begin{equation}
{\rm Prob}({\bf \rho}\mid {\bf D}, \tilde{\rho}) =
{\rm Prob}({\bf D}\mid {\bf \rho}, \tilde{\rho})\cdot
\frac{{\rm Prob}({\bf \rho}\mid \tilde{\rho})}
{{\rm Prob}({\bf D}\mid \tilde{\rho})}
\end{equation}
Here
$\tilde{\rho}$ is a prior set of the hypothesis
and ${\rm Prob}({\bf \rho}\mid \tilde{\rho})$ is the Bayesian prior probability.

 Maximizing the entropy of $\rho$ under $\langle \rho\rangle=\tilde{\rho}$,
$\langle\rho^2\rangle-\langle\rho\rangle^2=\tilde{\sigma}^2$ gives:
\begin{eqnarray}
{\rm Prob}(\rho\mid\tilde{\rho})&=& \frac{1}{\sqrt{2\pi\tilde{\sigma}}}
e^{-(\rho-\tilde{\rho})^2/2\tilde{\sigma}^2}\nonumber \\
{\rm Prob}(D \mid \rho, \tilde{\rho})&\propto & e^{-\chi^2/2}\nonumber \\
\chi_{prior}^2 &=&\sum_i \frac{(\rho_i-\tilde{\rho}_i)^2}{\tilde{\sigma}_i^2}\\
\Rightarrow \chi_{aug}^2 &=& \chi^2 + \chi_{prior}^2 
\end{eqnarray}
So the Bayesian theorem tells us that a single good inverse $\rho$ is to
minimize $\chi^2_{aug}$ \cite{Lepage}\cite{Morningstar}. 
Here the $\chi^2_{prior}$ plays the
role of the
additional minimizing term $\mathcal{B}(\rho)$ in Eq. (3) with $\lambda=1$.


\section{Systematic errors and multiplier ${\bf \lambda}$ }
In principle, minimizing $\mathcal{A}(\rho)$ and minimizing
$\mathcal{B}(\rho)$ would not necessarily give the same solution of $\rho$.
Hereby, multiplier vector ${\bf \lambda}$ is a bridge from the solution of
minimizing $\mathcal{A}$ to the solution of minimizing $\mathcal{B}$. With
$\lambda=1$ the constrained data modeling always introduces some systematic
bias which depends on how the input priors match the unknown
hypothesis. In some of the cases if we know the physics very well, we can
"teach the physics in fitting procedure". That means to input the
priors according to our knowledge to the hypothesis \cite{Lepage}.
However, in some of the cases we do not know the hypothesis well, only the
measured data set $D$ can tell us the information both of the priors
and hypothesis. We should
not simply maximize the probability ${\rm Prob}(D\mid\tilde{\rho})$ to
get the priors $\tilde{\rho}$, since that violates Bayesian theorem. Instead,
we can consider to use a subset of the measured data. 
In Bayesian statistics when some data set
$D_1$ comes along, and then some additional data set $D_2$ comes again,
the probability of $\rho$ in these two cases will be \cite{book}:
\begin{eqnarray}
{\rm Prob}(\rho\mid D_1, \tilde{\rho}) = {\rm Prob}(\rho\mid \tilde{\rho})
\cdot\frac{{\rm Prob}(D_1\mid \rho, \tilde{\rho})}
{{\rm Prob}(D_1\mid \tilde{\rho})} \nonumber \\
{\rm Prob}(\rho\mid D_2, \tilde{\rho}) = {\rm Prob}(\rho\mid \tilde{\rho})
\cdot\frac{{\rm Prob}(D_2\mid \rho, \tilde{\rho})}
{{\rm Prob}(D_2\mid \tilde{\rho})}
\end{eqnarray}
One can then prove the
estimate of $\rho$ probability in a enlarged data set:
\begin{eqnarray}
&&{\rm Prob}(\rho\mid D_2 D_1, \tilde{\rho}) \nonumber \\
&=& {\rm Prob}(\rho\mid \tilde{\rho}) \cdot
\frac{{\rm Prob}(D_2 D_1\mid \rho, \tilde{\rho})}
{{\rm Prob}(D_2 D_1\mid \tilde{\rho})},
\end{eqnarray}
which shows that we would have the same answer if all the data $D_1D_2$ had
been taken together. Furthermore, we can get the priors from data set $D_1$
then to fit $\rho$ from $D_1D_2$ taken together. That will not violate
the Bayesian theorem. So we can construct a real empirical Baye's method
to make data modeling. All the information comes from measured data set,
without any additional artificial bias. For example, we can give a 
programming
chart for the mass spectrum fitting such as in Fig. 1.

\begin{figure}[h]
\unitlength=0.4truecm
\begin{picture}(11,15)(0,-2.5)
\put(7.5,13){\oval(10,4)}
\put(3,13.8){\shortstack{\small{N=1, D$\geq$ 3, $\lambda_1=0$}}}
\put(3,12.0){\shortstack{\small{minimize $\chi^2_{aug}$}}}
\put(7.5,11){\vector(0,-1){0.9}}
\put(7.5,8.0){\oval(12.5,4)}
\put(1.8,8.8){\shortstack{\small{N$\Leftarrow$N+2, D$\Leftarrow$D+I, $\lambda_{N,N+1}=0$}}}
\put(3,7.0){\shortstack{\small{$\lambda_{i<N}=1$~~~~ minimize $\chi^2_{aug}$}}}
\put(7.5,6){\vector(0,-1){0.9}}
\put(7.5,4.0){\oval(5.5,2)}
\put(5,3.9){\small{\shortstack{? $N=N_{max}$}}}
\put(7.5,3){\vector(0,-1){2.5}}
\put(7.8,2.4){\shortstack{yes}}
\put(10.3,4){\vector(1, 0){4.5}}
\put(14.8,4){\vector(0, 1){3.6}}
\put(14.8,7.6){\vector(-1, 0){1.0}}
\put(11.2,4.2){\shortstack{no}}
\put(7.5,-1.5){\oval(12,4)}
\put(7.5,-3.5){\vector(0,-1){0.9}}
\put(2.5,-1.0){\small{\shortstack{$N=N_{max}$, D$\Leftarrow$D+M,
$\lambda_i=1$}}}
\put(2.5,-2.5){\small{\shortstack{$\lambda_{1,2}=0$~~~~
minimize $\chi^2_{aug}$}}}
\put(7.5,1.8){\vector(-1,0){4.5}}
\put(7.5,1.8){\circle*{0.3}}
\put(1.2,2.0){\oval(3.5,2.0)}
\put(0.1,2.3){\small{\shortstack{output}}}
\put(0.3,1.5){\small{\shortstack{priors}}}
\put(7.0,-5.0){\shortstack{ out}}
\end{picture}
\caption{{\small A programming chart of the empirical Baye's method. 
Where $I\neq 0$,
$M\neq 0$, $N$ is the number of the parameters in each fitting procedure, 
$N_{max}$
is the maximum number of the parameters we want to fit, $D$ is the number of 
the data points we use to fit the parameters, $\lambda_i$ is $i_{th}$ element
of the multiple vector ${\bf \lambda}$}}
\end{figure}
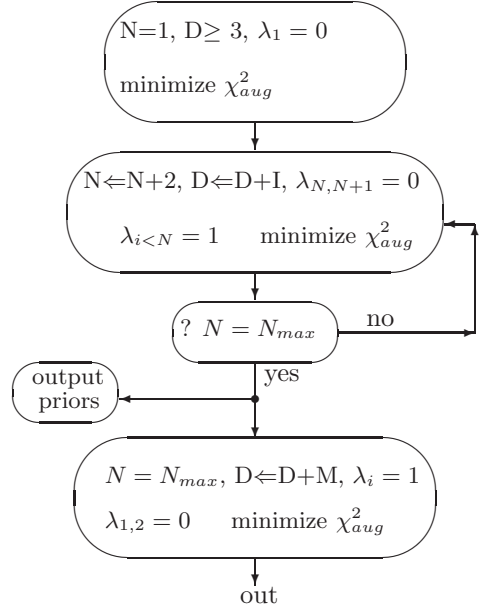
We test this empirical Baye's method on a $16^3\times 28$ lattice,
with $\beta=2.264$ Iwasaki gauge action. 
Quenched approximation with anti-periodic boundary in t-direction
is used. $f_\pi$ scale gives 
$a^{-1}=0.9775(47)$GeV.
The lowest pion mass is about $m_\pi\approx 179$MeV,
$m_\pi/m_\rho\approx 0.25$.
Empirical Baye's method works in such light quark area.
Fig. 2 shows the
$\lambda$ test at $m_\pi\approx 202$MeV. 
Where whole elements of vector $\lambda$ are equal to 1, 
except the first and second elements 
$\lambda_1,~\lambda_2$ are varying and $\lambda_1=\lambda_2$. This test
shows that the pion mass is stable, the priors were obtained from data subset 
matches full data set very well \cite{Terrence}.

An other advantage of this empirical Baye's method is that sometimes
we can observe
a stable excited state. Fig. 3 shows the nucleon mass and the mass of
the first
excited state. The mass ratio shows that the excited state is consistent with
the nucleon Roper state which lattice community has searched for years but
have not been successful before\cite{Frank}.
\[
\frac{m_{\rm exc}}{m_N}\sim 1.54\pm 0.17
\]
\newpage
\begin{figure}[ht]
\includegraphics{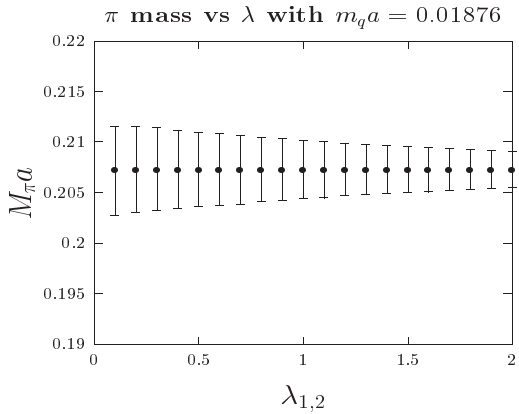}
\vspace{1.8in}
\caption{The $\lambda$ test for pion mass at $m_\pi\approx 202$MeV}
\end{figure}

\begin{figure}[ht]
\includegraphics{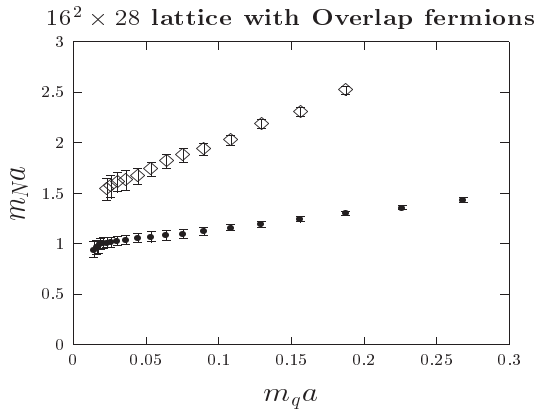}
\vspace{1.3in}
\caption{The Roper state as an excited state of the nucleon} 
\end{figure}
\section{Quenched chiral log fitting and the weak prior method}
 The quenched chiral log is a hard problem, since the general
minimum $\chi^2$ data modeling for the chiral log formula is singular 
\cite{CP-PACS}\cite{Terrence}.
The normal equation in the fitting procedure gives only 2 independent
parameters, the third one is not independent. 
\begin{equation}
m^2_{PS}  =  2 A m_q \{ 1-\delta[\ln(2Am_q/\Lambda^2_{\chi})+1]\}
 +O(m_q^2)
\end{equation}
An additional
matrix is really needed to lift the degeneracy in order 
to get the priors from data. 
An alternative way is to
use weak constrained data modeling to get the good priors. In this case 
we input
$\delta=0.20\pm 0.20,~\Lambda_\chi=1.1\pm 0.5$ as the weak priors. From a data
subset of
$m_\pi=179$MeV to $m_\pi=287$MeV to get the better priors. We then input this 
better priors into the full data set of $m_\pi=179$MeV to $m_\pi=312$MeV,
and obtain $\delta=0.26\pm 0.03$, $\Lambda = 1.1\pm 0.1$GeV. Fig. 4
shows the resultant fit of $(m_\pi a)^2$.

\begin{figure}[h]
\includegraphics{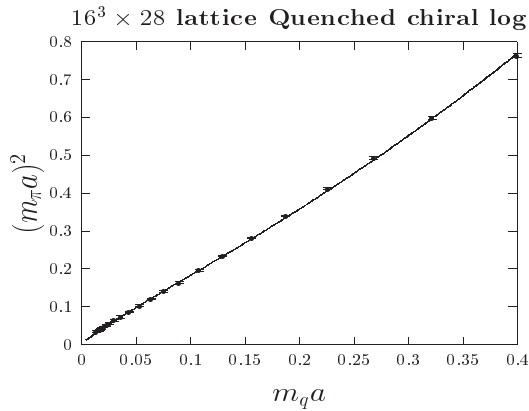}
\vspace{1.8in}
\caption{The weak constrained data modeling for the quenched chiral log} 
\end{figure}

Further work is on the way to improve the stability of the fitting program,
especially in the very light quark region, and try to automate the procedure.


\begin{thebibliography}{9}
\bibitem{book} S.J. Press, {\it Bayesian Statistics: Principles, Models and
Applications}, (Wiley, New York, 1989).
\bibitem{Lepage} G. P. Lepage, et al., Nucl.Phys.B(Proc. Suppl.) 106(2002)12
hep-lat/0110175. 
\bibitem{Morningstar} C. Morningstar, hep-lat/0112023.
\bibitem{Terrence} T. Draper et al., Proceeding of Lattice 2002 hep-lat/0208045,
S.J. Dong et.al., Nucl. Phys. B (Proc. Suppl.) 106 (2002) 275.
\bibitem{Frank} F.X. Lee et al., Proceeding of Lattice 2002 hep-lat/0208070,
Frank X. Lee et al., Nucl. Phys. B (Proc. Suppl) 106 (2002) 248.
\bibitem{CP-PACS} CP-PACS Collaboration, Phys. Rev. Lett. 84 (2000) 238.
\end{thebibliography}
\end{document}